\documentclass[conference]{IEEEtran}
\usepackage{hyperref}
\usepackage{url}
\usepackage{tabularx}
\IEEEoverridecommandlockouts
\usepackage{cite}
\usepackage{amsmath,amssymb,amsfonts}
\usepackage{algorithm}
\usepackage{float}
\usepackage{amsmath}
\usepackage{graphicx}
\usepackage{textcomp}
\usepackage{algpseudocode}
\usepackage[utf8]{inputenc}
\usepackage{xcolor}
\usepackage{acronym}
\def\BibTeX{{\rm B\kern-.05em{\sc i\kern-.025em b}\kern-.08em
    T\kern-.1667em\lower.7ex\hbox{E}\kern-.125emX}}
\begin{document}

\acrodef{OFDMA}{Orthogonal Frequency Division Multiple Access}
\acrodef{OFDM}{Orthogonal Frequency Division Multiplexing}
\acrodef{RU}{Resource Unit}
\acrodef{STA}{Station}
\acrodef{MU-MIMO}{Multi-User Multiple Input Multiple Output}
\acrodef{EDCA}{Enhanced Distributed Channel Access}
\acrodef{AP}{Access Point}
\acrodef{TSN}{Time-Sensitive Networking}
\acrodef{IIoT}{Industrial Internet of Things}
\acrodef{NG-RAN}{Next Generation Radio Access Network}

\title{Exploratory Analysis of Wi-Fi 6 Dynamic RU Sharing in Small-Scale Network Scenarios}

\author{\IEEEauthorblockN{Sai Geetha Mada}
\IEEEauthorblockA{\textit{Department of Computer Science (INF)} \\
\textit{Friedrich-Alexander-Universität Erlangen-Nürnberg}\\
}
\and
\IEEEauthorblockN{Rute C Sofia}
\IEEEauthorblockA{fortiss GmbH, Munich
}
\and
\IEEEauthorblockN{Anna Baron}
\IEEEauthorblockA{\textit{Department of Computer Science (INF)} \\
\textit{Friedrich-Alexander-Universität Erlangen-Nürnberg}\\
}
\and
\IEEEauthorblockN{Luigi Martino}
\IEEEauthorblockA{\textit{Department of Computer Science (INF)} \\
\textit{Friedrich-Alexander-Universität Erlangen-Nürnberg}\\
}
}

\maketitle

\begin{abstract}
This paper investigates dynamic Resource Unit (RU) allocation strategies for Wi-Fi~6 (IEEE 802.11ax) networks integrated with Time-Sensitive Networking (TSN), targeting the limitations of static RU scheduling under dynamic traffic conditions. We propose a dynamic RU allocation algorithm that maps TSN traffic classes to Wi-Fi~6 Quality of Service (QoS) mechanisms, including Enhanced Distributed Channel Access (EDCA) and aligns TSN control with Ethernet-based TSN domains. The proposed solution is evaluated using the ns-3 DetNetWiFi framework developed by fortiss, focusing on time-sensitive traffic. Simulation results demonstrate improved network efficiency with reductions in latency, jitter, and packet loss compared to static RU allocation schemes. These findings highlight the potential of dynamic RU allocation to support deterministic communication requirements in Wi-Fi~6-based TSN deployments and to enhance the reliability of hybrid industrial networks.
\end{abstract}


\section{Introduction}
The proliferation of latency-critical and high-reliability applications in industrial automation, robotics, and advanced manufacturing has intensified the demand for wireless networks capable of delivering deterministic performance. Wi-Fi~6 (IEEE 802.11ax) addresses part of this challenge through \textit{Orthogonal Frequency Division Multiple Access (OFDMA)}, which enhances spectral efficiency and supports high client density. However, the default static allocation of \textit{Resource Units (RUs}) in Wi-Fi~6 is insufficient in dynamic traffic environments, often resulting in increased latency, jitter, and underuse channel capacity. These limitations are adverse to time-sensitive and mission-critical applications~\cite{mozaffariahrar2022wifi6}.

\textit{Time-Sensitive Networking (TSN}) offers deterministic guarantees such as bounded latency, zero packet loss, low jitter. TSN traditionally operates over wired Ethernet infrastructures. Extending TSN capabilities to wireless domains has emerged as a promising direction for industrial connectivity, yet doing so requires precise coordination of QoS policies, scheduling mechanisms, and resource allocation strategies. 

Integrating TSN with Wi-Fi~6 introduces challenges related to harmonizing traffic priorities and ensuring that TSN service guarantees are preserved in an OFDMA-based wireless environment~\cite{mozaffariahrar2022wifi6}. In particular, effective alignment between TSN traffic classes and Wi-Fi~6 QoS mechanisms such as \textit{Enhanced Distributed Channel Access (EDCA)} is essential to maintain determinism across heterogeneous segments.

To address these challenges, this work proposes a dynamic \textit{Resource Unit (RU)} allocation strategy designed to improve Wi-Fi~6 performance for TSN-enabled traffic. The proposed approach adapts RU assignments based on real-time traffic variations and maps TSN traffic classes to Wi-Fi~6 EDCA categories, enabling more predictable and efficient resource utilization. 

The proposed algorithm is implemented and evaluated via simulations carried out with the \textit{network simulator version 3 (ns-3)} plus the ns-3 DetNetWiFi simulation framework~\footnote{https://www.fortiss.org/en/results/software/detnetwifi-network-simulator-v30-framework}. 

The simulations are initial and exploratory. The scenarios are intentionally constrained to allow controlled observation of system behavior, and the results should be interpreted as exploratory rather than definitive. The simulations examine key performance indicators including latency, jitter, packet loss, and channel utilization under varying traffic conditions. 

Additionally, interoperability with a 5G core is explored to ensure compatibility with future industrial communication architectures and to assess the scalability of the approach in broader hybrid deployments~\cite{satka2023experimental}. 

The remainder of this paper is organized as follows. 
Section~\ref{relatedwork} reviews related work on dynamic RU allocation, wireless TSN integration, and QoS mapping mechanisms. 
Section~\ref{sec:background} introduces the main technological enablers, namely Wi-Fi~6, TSN, 5G, and existing RU allocation strategies in OFDMA-based WLANs. 
Section~\ref{sec:Wi-Fi-Ethernet} presents the first ns-3 case study, detailing the TSN-aware Wi-Fi/Ethernet modelling and QoS mapping between EDCA and TAS. 
Section~\ref{sec:Wi-Fi-5G} extends this modelling to a converged TSN Wi-Fi~6 over 5G core scenario. 
Section~\ref{sec:tsn-ru-model} specifies the proposed TSN-aware dynamic RU allocation and scheduling model for Wi-Fi~6. 
Section~\ref{sec:evaluation} reports the performance evaluation for two exploratory case studies (C1 and C2), focusing on QoS behaviour and dynamic RU sharing under varying load conditions. 
Finally, Section~\ref{sec:conclusion} concludes the paper by summarizing the main findings and outlining directions for future, more extensive experimental studies.

\section{Related Work}
\label{relatedwork}
Efficient RU allocation is fundamental to enabling the concurrent multi-user transmission capabilities introduced by OFDMA in Wi-Fi 6. 
With the increase of TSN Wi-Fi complementing fixed or cellular cores, the limitations of traditional RU scheduling becomes increasingly evident. 

Traditionally, the existing static RU allocation supported by vendors provides wireless \textit{Access Points (APs)} with the capability to assign fixed resources to wireless stations based on predetermined rules, such as predefined priority levels or fixed bandwidth slices. Its simplicity and low implementation overhead have made it a common baseline in many Wi-Fi deployments. Prior studies, including \cite{khorov2018tutorial} and \cite{bellalta2020ofdma}, show that static allocation performs reasonably well in networks with stable and predictable traffic patterns.
However, the method struggles in environments with variable traffic demands. IIoT environments are characterized by heterogeneous, bursty, and often mission-critical traffic, which require a dynamic RU allocation. Static RU allocation may result in bandwidth underutilization during low-load periods and excessive latency during high-load or priority-sensitive operations. Moreover, static allocation lacks mechanisms to prioritize time-sensitive or deterministic flows, resulting in increased jitter and reduced reliability.
Round-robin scheduling assigns RUs in a cyclic, equal-turn fashion, offering fairness across stations. Each device receives a transmission opportunity regardless of traffic type or urgency. While this ensures fair access and reduces implementation complexity, research such as \cite{zhang2021roundrobin} demonstrates that the approach is poorly suited to heterogeneous traffic profiles.
In particular, round-robin scheduling does not distinguish between critical and non-critical traffic. Time-sensitive streams—such as TSN control messages, voice, or emergency communication—are treated the same as background updates or low-priority telemetry. As a result, deterministic performance guarantees cannot be met, making round-robin unsuitable for hybrid TSN–Wi-Fi 6 networks where QoS differentiation is essential.

To address the shortcomings of traditional techniques, recent research increasingly emphasizes dynamic RU allocation. These adaptive approaches leverage real-time network state information such as buffer occupancy, channel conditions, TSN priority classes, and application QoS parameters to guide resource scheduling decisions. Studies such as \cite{cena2022dynamicscheduling} show that dynamic algorithms can significantly improve throughput, latency, and jitter performance compared to static or round-robin methods.

\section{Enablers of TSN Wired–Wireless Convergence}
\label{sec:background}
This section summarizes the key technologies underpinning this work, namely Wi-Fi~6, TSN, 5G, and existing RU allocation strategies in \ac{OFDMA}-based WLANs, as well as their joint integration in hybrid wired–wireless networks. In this paper, we use ‘RU’ as a generic schedulable time–frequency block across both domains; in 5G this corresponds to PRBs/scheduler grants, while in Wi-Fi 6 it corresponds to OFDMA RUs.

\subsection{Wi-Fi~6 and OFDMA-Based Multi-User Access}
Wi-Fi~6 (IEEE 802.11ax) extends the functionality of Wi-Fi~5 (802.11ac) by introducing multi-user features designed for dense deployments and heterogeneous traffic. While Wi-Fi~5 relies on \ac{OFDM} with single-user transmissions per channel, Wi-Fi~6 adopts \ac{OFDMA}, which divides the channel into smaller \acp{RU} assigned to multiple \acp{STA} in parallel. This reduces contention, improves spectral efficiency, and lowers latency, making Wi-Fi~6 suitable for real-time and IoT traffic. In addition, Wi-Fi~6 extends \ac{MU-MIMO} to both uplink and downlink and increases the number of supported spatial streams, further enhancing capacity in high-density environments. Features such as Target Wake Time (TWT) improve energy efficiency, particularly for battery-powered IoT devices.

\subsection{EDCA and QoS in Wi-Fi~6}

\textit{Enhanced Distributed Channel Access (\ac{EDCA})} is the primary QoS mechanism in Wi-Fi, including Wi-Fi~6. It classifies traffic into Access Categories (AC\_VO, AC\_VI, AC\_BE, AC\_BK) with differentiated contention parameters (CW, AIFS, TXOP) so that high-priority flows (e.g., voice, critical control messages) obtain faster medium access than best-effort or background traffic. In converged wired–wireless networks, EDCA at the \ac{AP} plays a key role in aligning wireless QoS with wired QoS policies, especially when integrating with TSN-enabled Ethernet segments.

\subsection{Time-Sensitive Networking (TSN)}

\ac{TSN} is a family of IEEE 802.1 standards that provides deterministic communication over Ethernet with guarantees on latency, jitter, and reliability. Core mechanisms include time-aware shaping (IEEE 802.1Qbv), per-stream filtering and policing, traffic shaping, and tight clock synchronization (IEEE 802.1AS). TSN is widely adopted in industrial automation, automotive, healthcare, and smart grid applications, where control loops and safety-critical traffic require bounded delay and high reliability. TSN’s ability to integrate with existing Ethernet infrastructure makes it a natural foundation for next-generation \ac{IIoT} networks, but extending its guarantees over wireless segments remains challenging.

\subsection{5G Support for Deterministic Communication}

5G introduces architectural and functional advances that align closely with TSN requirements. \textit{Ultra-Reliable Low-Latency Communication (URLLC)} targets time-sensitive applications, while a Service-Based Architecture (SBA) and network slicing enable the creation of dedicated logical networks with tailored QoS. Support for time-sensitive networking is provided through 5G–TSN interworking mechanisms, including the \textit{TSN Application Function (TSN AF)} and the exchange of \textit{TSN Assistance Information (TSCAI)}, rather than through a standalone TSN feature. Precise time synchronization (e.g., gPTP/IEEE 802.1AS integration) and \textit{Multi-access Edge Computing (MEC)} further reduced latency and improved determinism. The 5G QoS framework allows mapping of TSN flows to dedicated QoS profiles, and TSN-aware bridging functions in the 5G core and \ac{NG-RAN} extend deterministic behavior across wired and wireless domains.

\subsection{RU Allocation in Wi-Fi~6}

RU allocation is central to the performance of OFDMA-based Wi-Fi~6 systems. Static RU allocation, where fixed RUs are assigned using predefined rules, is simple but rigid: it cannot adapt to varying traffic loads and does not adequately prioritize time-sensitive flows, leading to suboptimal utilization and increased latency in dynamic or industrial environments. Round-robin scheduling offers fairness by cyclically assigning RUs to stations, but it is oblivious to QoS requirements and traffic heterogeneity, making it unsuitable for TSN-integrated scenarios that require strict prioritization and deterministic behavior. Recent work has therefore begun to explore dynamic RU allocation strategies that use real-time traffic information, QoS requirements, and sometimes machine learning to adapt RU assignments. Such approaches aim to better support latency-critical traffic while maintaining high spectral efficiency in hybrid TSN–Wi-Fi~6 networks.

\subsection{Integration of TSN, Wi-Fi~6, and 5G}
The joint integration of TSN, Wi-Fi 6, and 5G is a key step toward end-to-end deterministic communication across heterogeneous infrastructures, particularly in industrial and enterprise environments where a single wireless technology is insufficient. While 5G provides wide-area coverage, mobility support, URLLC, network slicing, and edge computing, Wi-Fi 6 remains attractive for local and indoor deployments due to its cost efficiency, high device density support, and tight integration with existing Ethernet-based TSN infrastructures.

TSN over Ethernet provides precise scheduling, shaping, and synchronization in wired segments; Wi-Fi 6 contributes flexible multi-user access and QoS-aware scheduling in wireless LANs; and 5G extends deterministic communication across wide-area and mobile domains. Achieving seamless determinism across these domains requires consistent QoS mapping (e.g., TSN classes to EDCA ACs and 5QI), tight time synchronization, and coordinated scheduling, as well as scalable resource allocation mechanisms such as dynamic RU allocation.

\subsection{Dynamic Resource Allocation and QoS in Wi-Fi 6 and TSN Integration}
Wi-Fi 6 introduces several features that can support TSN integration, particularly in the areas of dynamic resource allocation and QoS management. In former work, \textit{Time-aware Shaping (TAS)} has been considered based on \textit{Target-Wake Time (TWT)} to support deterministic behaviour in the scheduling, ensuring the deterministic transmission of time-sensitive traffic \cite{schneider2022proposal}. 

Additionally, traffic scheduling is significantly improved through features like OFDMA and MU-MIMO, which enable more efficient medium access control, allowing deterministic scheduling of time-sensitive traffic. Furthermore, dynamic resource allocation in Wi-Fi 6 utilizes adaptive scheduling mechanisms that help manage varying network loads while maintaining stringent QoS requirements. These features collectively enhance the ability of Wi-Fi 6 to support TSN applications by improving reliability, reducing latency, and optimizing resource allocation.

\begin{figure}[!h]
    \centering
    \includegraphics[width=1\columnwidth]{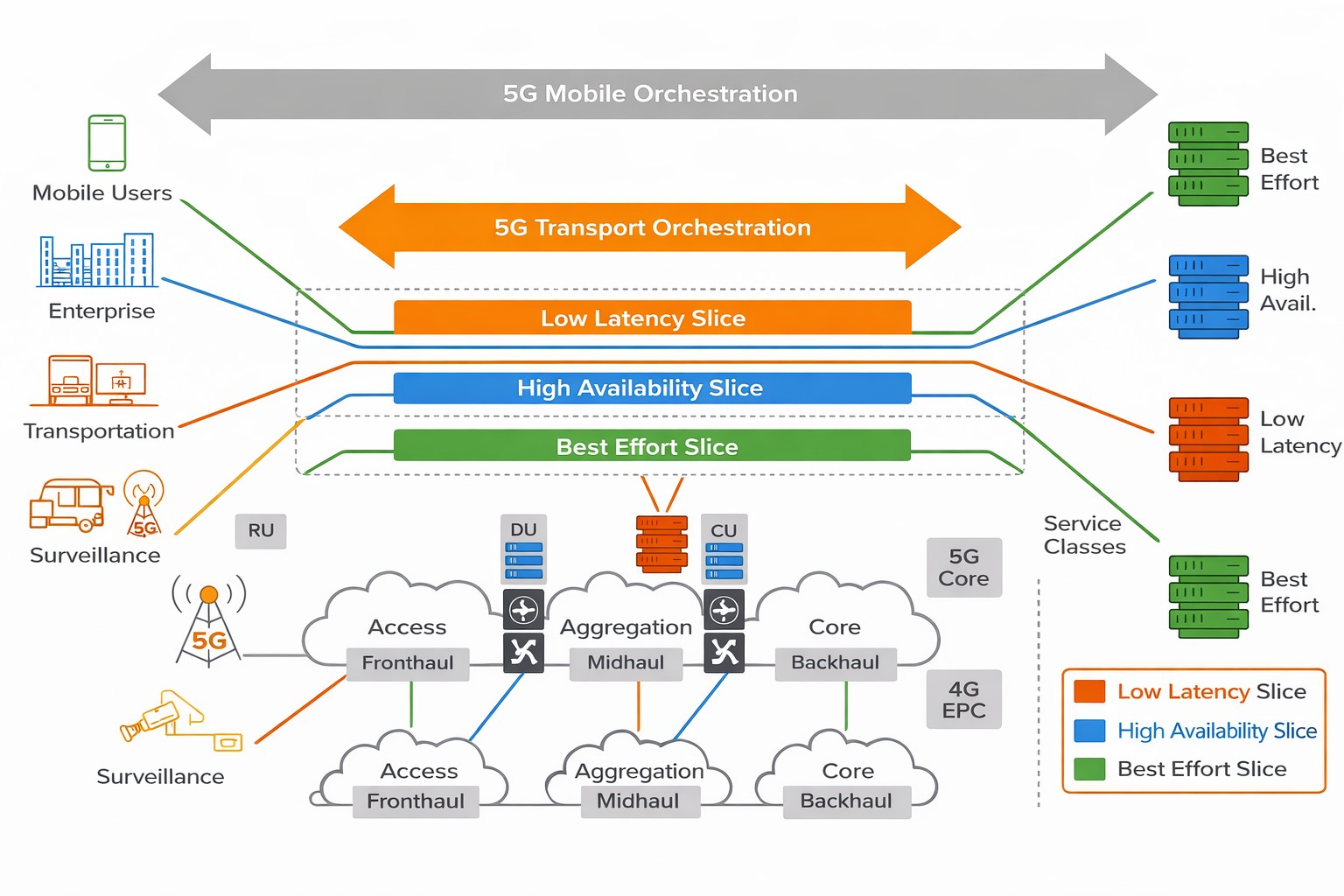}
    \caption{ 5G Network Slicing. Diagram adapted from Infinera’s blog on 5G
slicing and network optimization \cite{infinera_5gslicing_2025}.}
    \label{fig:slicing}
\end{figure}

Figure~\ref{fig:slicing} illustrates a simplified system topology representing the hybrid Wi-Fi/Ethernet TSN environment considered in this study.

The integration of 5G networks with TSN is being actively standardized in 3GPP Release 16 and 17 to enhance industrial automation by ensuring reliable and deterministic communication. One key aspect of this convergence is network slicing which is showed in Fig. 1, which allows 5G to create virtual networks customized to meet specific QoS requirements. This feature facilitates efficient resource allocation for TSN traffic, ensuring that time-sensitive applications receive the necessary network performance. Additionally, adaptive scheduling plays a crucial role in optimizing resource management within 5G networks. By dynamically adjusting resource allocation based on real-time network conditions and QoS demands, adaptive scheduling helps maintain the stringent latency and reliability requirements essential for TSN-enabled applications \cite{b8}. Another critical factor in achieving seamless integration is QoS mapping, which ensures that QoS parameters from TSN and 5G systems are aligned to maintain consistent service quality. Various methodologies have been explored to effectively map these parameters, addressing the challenge of interoperability between wired and wireless TSN networks \cite{b9}.

\subsection{Current Gaps and Challenges}
Integrating TSN with Wi-Fi 6 and 5G presents several challenges that must be addressed to achieve reliable and deterministic communication. Standardization remains a key issue, as TSN features need further refinement to ensure interoperability over wireless networks \cite{shi2021evaluating}. Time synchronization is another major challenge, as maintaining precision in wireless environments is difficult due to varying link quality and interference \cite{shi2021evaluating}. Additionally, resource management requires advanced scheduling techniques to meet the stringent latency and reliability requirements of TSN traffic in 5G networks \cite{nasrallah2018ultra}.

To bridge these gaps, innovative strategies are being developed, such as dynamic RU allocation in Wi-Fi 6 based on TSN QoS classes to achieve deterministic scheduling for TSN traffic. Simulation frameworks like DetNetWiFi~\footnote{https://www.fortiss.org/en/results/software/detnetwifi-network-simulator-v30-framework}, built on the ns-3 network simulator, are used to model and evaluate these strategies, enabling the development and testing of mechanisms that support critical applications in IIoT settings. 

\section{TSN Wi-Fi/Ethernet Modelling with ns-3}
\label{sec:Wi-Fi-Ethernet}

This case study models, in ns-3, a converged Wi-Fi–Ethernet architecture that supports both time-sensitive and best-effort traffic with consistent QoS guarantees across wired and wireless segments. The ns-3 DetNetWiFi framework has been extended to capture the main functional blocks required for TSN-aware operation in a mixed Wi-Fi/Ethernet environment. The system model is structured around two key aspects:
\begin{enumerate}
    \item Integration of Wi-Fi~6 and TSN-enabled Ethernet with aligned QoS mechanisms, using EDCA in the wireless segment and TAS in the wired segment.
    \item Design and implementation of a TSN-aware dynamic \textit{Resource Unit (RU)} allocation algorithm for Wi-Fi~6.
\end{enumerate}

The hybrid topology consists of a Wi-Fi~6 access segment and a TSN-capable Ethernet backbone, interconnected by an AP interconnecting the two IP subnets: one for the Wi-Fi region and one for the TSN/Ethernet region. All elements are instantiated and interconnected within the ns-3 scenario, enabling a reproducible and configurable case study.

\subsection{Access Point}
In the model, the AP is instantiated using the \textit{ns3::ApWifiMac} class and configured as an IEEE~802.11ax (Wi-Fi~6) access point. This choice reflects the need to represent modern WLAN capabilities, including high spectral efficiency, dense device deployments, and native QoS support. The MAC layer is enabled with QoS extensions to differentiate traffic classes\footnote{The complete implementation code for this setup can be accessed at \url{https://git.fortiss.org/iiot_external/ns-3-detnetwifi}}.

Within the case study, the AP is modelled as the traffic orchestrator of the wireless domain. It applies EDCA-based prioritization to ensure that time-critical data is expedited while routine traffic is deferred when needed. Whenever Wi-Fi stations contend for the medium, the AP drives access control via EDCA, mapping each flow to one of the four EDCA \textit{Access Categories (ACs)}: Voice (\textit{AC\_VO}), Video (\textit{AC\_VI}), Best Effort (\textit{AC\_BE}), and Background (\textit{AC\_BK}). Real-time and delay-sensitive traffic is directed to higher-priority ACs, whereas non-urgent data is associated with lower-priority categories.

To realize these distinctions, the AP is parameterized through the standard QoS parameters: \textit{Contention Window (CW)}, \textit{Arbitration Interframe Space Number (AIFSN)}, and \textit{Transmission Opportunity (TXOP)}. In the model, smaller CW values and shorter AIFSN intervals are configured for high-priority ACs, enabling faster channel access, while TXOP limits bound the maximum transmission duration per opportunity.

The TSN-aware RU allocation algorithm is also modeled as part of the AP logic. It ensures that the priority levels assigned in the Wi-Fi domain (via EDCA parameters and access categories) are coherently translated when transitioning to TAS-based scheduling on the Ethernet side, maintaining QoS consistency and minimizing additional latency at the boundary.

\subsection{Ethernet Switch}
The Ethernet switch is modeled as the backbone element of the wired TSN domain, interconnecting the AP with wired TSN/Ethernet devices. In addition to basic packet forwarding, the switch represents deterministic traffic scheduling capabilities through a TAS driven by Gate Control Lists (GCLs), providing bounded latency (minimum and maximum), near-zero packet loss for critical streams, and low jitter.

The switch and its attached links are instantiated using the \textit{ns3::Csma} model, which captures high-speed, low-latency Ethernet behaviour. The following configuration parameters are used:
\begin{itemize}
    \item \textbf{Data Rate:} 1~Gbps, to represent high-speed Ethernet segments.
    \item \textbf{Propagation Delay:} 6.56~\textmu s, approximating typical physical-layer delays.
\end{itemize}

\subsection{TAS and GCL Modelling}

In the case study, TAS is modelled by explicitly configuring \textit{Gate Control Lists (GCLs)} at the switch. Each GCL entry specifies the exact time intervals during which traffic from a given queue is allowed to transmit or must remain blocked.

Traffic classification is performed at the Ethernet switch via IEEE~802.1Q VLAN tagging, where Priority Code Points (PCPs) filed within the VLAN tag is used to encode packet priority. Upon ingress to the switch, the PCP value is extracted from the VLAN tag and mapped to a corresponding TAS \textit{Traffic Class (TC)}. GCLs operate on these traffic classes by controlling the open and closed states of their associated transmission queues.

Priority classification in the switch is defined as:
\begin{itemize}
    \item Emergency traffic (highest priority) $\rightarrow$ PCP~7,
    \item Voice traffic $\rightarrow$ PCP~6,
    \item Video traffic $\rightarrow$ PCP~5,
    \item Best-effort traffic $\rightarrow$ PCP~3,
    \item Background traffic (lowest priority) $\rightarrow$ PCP~1.
\end{itemize}

GCLs determine when each traffic queue is allowed to transmit, ensuring predictable bandwidth allocation. In the model, each gate alternates between open and closed states at fixed intervals, providing dedicated windows for time-sensitive traffic and opportunistic access for lower-priority queues.

The TAS cycle is represented as a sequence of time slots, each corresponding to an entry in the GCL (rf. to Table~\ref{tab:traffic_class}). In the reference configuration, the TAS cycle is set to 100~ms and divided into four equal 25~ms slots, with a Boolean GCL pattern \texttt{true, false, true, false}. During the first and third slots, gates for high-priority queues are opened; during the second and fourth slots, gates remain closed, and no traffic is transmitted from those queues. This periodic gating pattern yields deterministic access for critical flows while restricting transmissions in other periods.

\begin{table}[h!]
    \centering
    \caption{TAS Traffic class priorities and their active slots and cycle time.}
    \label{tab:traffic_class}
    \begin{tabular}{|l|l|l|l|}
        \hline
        \textbf{Traffic Class} & \textbf{Priority} & \textbf{Active Slots} & \textbf{Cycle Time} \\ \hline
        Emergency & 7 & 2/4 & 100 ms \\ \hline
        Voice & 6 & 1/4 & 100 ms \\ \hline
        Video & 5 & 1/4 & 100 ms \\ \hline
        Best Effort & 3 & Opportunistic & 100 ms \\ \hline
        Background & 1 & Opportunistic & 100 ms \\ \hline
    \end{tabular}
\end{table}

The choice of 25~ms slots within a 100~ms cycle is deliberate, balancing low-latency requirements, efficient bandwidth usage, and fair access to resources. In particular, high-priority traffic such as emergency signaling or time-sensitive control loops benefit from frequent transmission opportunities, keeping queueing delays bounded.

\subsection{Mapping Between Wi-Fi and Ethernet QoS Priorities}

To provide an end-to-end TSN-compliant case study, the model includes explicit mapping rules between Wi-Fi (EDCA) priorities and Ethernet (TAS) traffic classes. Table~\ref{tab:wifi_ethernet_mapping} summarizes the mapping strategy used in the scenario. It ensures that traffic retains a consistent level of service regardless of whether it is carried over Wi-Fi or Ethernet.

\begin{table}[h]
    \centering
    \caption{Mapping of Wi-Fi Access Categories to Ethernet Traffic Classes and PCP Values.}
    \label{tab:wifi_ethernet_mapping}
    \resizebox{\columnwidth}{!}{
    \begin{tabular}{|l|l|c|}
    \hline
    \textbf{Wi-Fi Access Category (EDCA)} & \textbf{Ethernet Traffic Class (TAS Priority)} & \textbf{PCP Value (802.1p)} \\ \hline
    Voice (AC\_VO) & Emergency (Priority 0) & 7 \\ \hline
    Video (AC\_VI) & Voice (Priority 1) & 5 \\ \hline
    Best Effort (AC\_BE) & Best Effort (Priority 3) & 3 \\ \hline
    Background (AC\_BK) & Background (Priority 4) & 1 \\ \hline
    \end{tabular}
    }
\end{table}

\begin{figure*}[h]
    \centering
    \includegraphics[width=1\linewidth]{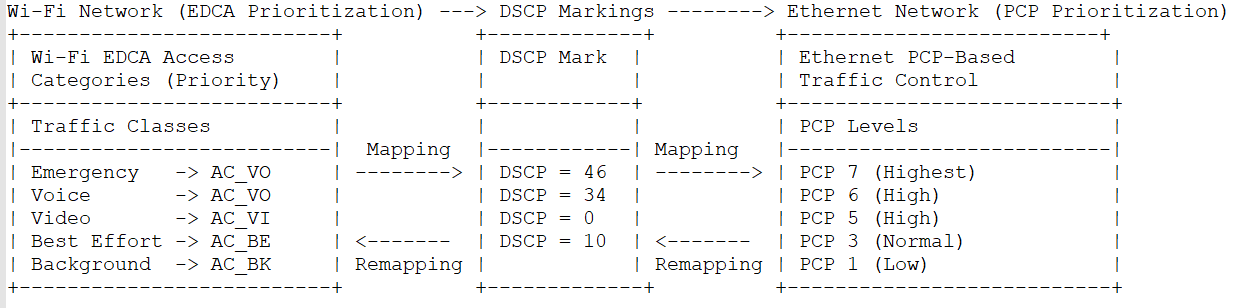}
    \caption{Mapping and remapping of QoS priorities between Wi-Fi and Ethernet.}
    \label{fig:mapping}
\end{figure*}

Figure~\ref{fig:mapping} illustrates how these mappings are applied to traffic crossing the Wi-Fi/Ethernet boundary. By maintaining consistent class assignments and priorities, the model enables deterministic communication behavior and stable QoS across the hybrid network.

QoS mechanisms are modeled across both segments to preserve end-to-end prioritiation.

On the Wi-Fi 6 region, the ns-3 model uses EDCA to prioritize traffic based on ACs. For example:
\begin{itemize}
    \item Emergency traffic is mapped to \textit{AC\_VO} with minimal CW and AIFSN values (rcf.~Table~\ref{tab:ACs}) to reduce access delay.
    \item Video traffic is mapped to \textit{AC\_VI}, which receives better service than Best Effort but lower priority than Voice.
\end{itemize}

\begin{table}[h!]
\centering
\caption{Access Categories and Parameters.}
\label{tab:ACs}
\begin{tabular}{|l|c|c|c|c|}
\hline
\textbf{AC} & \textbf{MinCw} & \textbf{MaxCw} & \textbf{AIFS} & \textbf{TXOP} \\ 
\hline
AC\_VO (Voice) & 1 & 3 & 1 & 2048 \\ 
\hline
AC\_VI (Video) & 7 & 15 & 2 & 3008 \\ 
\hline
AC\_BE (Best Effort) & 15 & 1023 & 3 & -- \\ 
\hline
AC\_BK (Background) & 15 & 1023 & 7 & -- \\ 
\hline
\end{tabular}
\end{table}

In the case study, the AP and Ethernet switch jointly manage traffic through EDCA and TAS respectively, while the mapping and shaping logic resides in the AP. EDCA provides dynamic prioritization on Wi-Fi, whereas TAS offers deterministic scheduling on Ethernet. 

In the \textbf{uplink} direction (Wi-Fi to Ethernet), packets from Wi-Fi station nodes are first prioritized at the AP via EDCA Figure ~\ref{fig:scenario2}, then forwarded to the Ethernet switch where TAS scheduling enforces deterministic treatment for high-priority traffic.

In the \textbf{downlink} direction (Ethernet to Wi-Fi), packets generated by Ethernet devices are first scheduled by TAS at the switch, then delivered to the AP, where EDCA-based AC mapping ensures that time-sensitive flows (e.g., voice and emergency traffic) experience minimal additional delay.

To preserve QoS semantics across domains, the AP also acts as a translation point for priority labels. In the Wi-Fi segment, traffic is categorized into ACs via EDCA. 
Upon reception of IP packets, the AP marks packets with \textit{Differentiated Services Code Point (DSCP)} values and, before forwarding traffic onto Ethernet segment, encapsulates packets into IEEE 802.1Q VLAN tagged Ethernet frames with corresponding PCP values. Ethernet TSN switches operate at Layer 2 and rely exclusively on the PCP field for traffic classification and TAS schedulling. For instance, emergency traffic in \textit{AC\_VO} is marked with DSCP~48 and mapped to PCP~7; video traffic in \textit{AC\_VI} is marked with DSCP~34 and mapped to PCP~7. Lower-priority ACs are associated with smaller DSCP values and lower PCP priorities. The reverse mapping, from PCP/DSCP back to EDCA ACs, is applied in the downlink direction, as illustrated in Figure~\ref{fig:mapping}.

\section{TSN Wi-Fi 6 Over a 5G Core: Convergence Modelling in ns-3}
\label{sec:Wi-Fi-5G}

The second case study (C2) addresses ns-3 modeling to support a converged TSN Wi-Fi 6 region to 5G core integration. The model combines TSN mechanisms (TAS, VLAN/PCP marking, GCL-based scheduling) with Wi-Fi~6 OFDMA and EDCA prioritization and 5G Core QoS flows (5QI, GBR/MBR, PDB). All these components are instantiated using ns-3 modules and extended interfaces as explained in Section V-A.

To orchestrate heterogeneous traffic, the implementation uses ns-3’s 5G-LENA module \footnote{https://5g-lena.cttc.es/}, extended to interact with Wi-Fi~6 and TSN Ethernet elements. The 5G Core is represented by the \textit{PointToPointEpcHelper}, which handles session management and end-to-end QoS enforcement, while a convergence layer preserves TSN traffic class semantics across Wi-Fi~6 and 5G RAN boundaries.

\subsection{Wi-Fi 6 Regions and TSN–5G Interconnection}

The TSN Wi-Fi~6 region is interconnected to the 5G Core via a TSN-enabled switch (ETH1), which models the bridging function between the Wi-Fi~6 AP and the 5G gNB. ETH1 is responsible for preserving TSN priorities as traffic flows between the two wireless domains and the core.

In the ns-3 model, ETH1 extends the TSN-aware Ethernet segment into the 5G Core by applying VLAN PCP and DSCP tags that are consistent with TSN class definitions. These markings are then mapped to the appropriate 5QI profiles inside the 5G Core, allowing high-priority TSN streams to maintain precedence even under varying load conditions.

\begin{figure}[h!]
    \centering
    \includegraphics[width=0.75\linewidth]{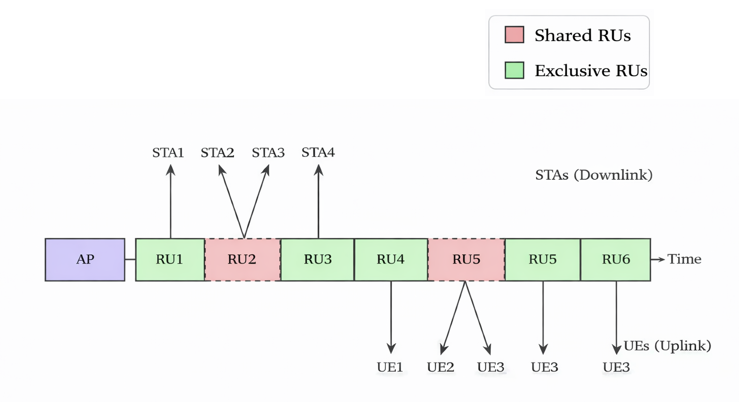}
    \caption{Dynamic RU allocation across TSN Wi-Fi~6 and 5G domains.}
    \label{fig:DRA}
\end{figure}

Figure~\ref{fig:DRA} shows the conceptual interconnection, highlighting how traffic is forwarded between the Wi-Fi~6 AP, the gNB, and the core network. ETH1 and its core-facing interface are key elements in preserving QoS consistency and ensuring efficient resource utilization in this hybrid TSN Wi-Fi 6 to 5G network.\footnote{The complete implementation code for this setup can be accessed at \url{https://git.fortiss.org/iiot_external/ns-3-detnetwifi}}.

Support for TAS from the TSN Wi-Fi~6 region into the 5G Core is captured through the coordination of GCLs, 5G Core QoS mechanisms, and Wi-Fi~6 prioritization features.

Within the 5G Core, traffic is classified using 5QI profiles, which encode priority, latency tolerance, and reliability requirements. \textit{Guaranteed Bit Rate (GBR)} flows reserve bandwidth for critical TSN applications (e.g., motion control), while \textit{Maximum Bit Rate (MBR)} parameters cap the usage of non-critical flows. The Packet Delay Budget (PDB) defines the maximum one way delay allowed for a QoS flow within the 5G system and directly influences scheduler prioritization and queue management. By configuring PDB values in accordance with TSN timing requirements, the model enables delay-aware schedulling and realistic end-to-end latency evaluation.

Dynamic resource allocation is modelled through the \textit{AllocateRusBasedOnFlows()} logic, which assigns RUs according to instantaneous demand and QoS requirements, thereby enabling deterministic latency for time-critical streams. QoS flow mapping then links 5G bearer services to TAS traffic classes, ensuring consistent priority semantics across protocol layers.

In the TSN switch, GCLs enforce transmission windows that minimize jitter and guarantee bounded delivery times. The TAS cycle is configured to align with 5G scheduling periodicity, so that deterministic flows remain synchronized as they cross wired–wireless boundaries. VLAN tagging using PCP and DSCP ensures that high-priority TSN packets are mapped to appropriate GBR bearers in the 5G Core, maintaining end-to-end priority.

On the Wi-Fi~6 side, EDCA differentiates traffic into access categories, ensuring that TSN flows are scheduled ahead of best-effort data. CWmin/CWmax and TxOp limits parameters are configured according to Table III are tuned to reduce backoff delays and maximum channel occupancy, thereby reducing access delay and providing prioritized airtime for TSN flows over best-effort traffic.

The 5G gNB uses the \textit{RrFfMacScheduler} to allocate radio resources, cycling RUs across UEs while supporting TSN-aligned GBR flows mapped through specific 5QI profiles. Key functions in the model are:
\begin{itemize}
    \item mapping TSN TC/PCP/DSCP values to 5QI classes,
    \item enforcing GBR/MBR profiles to protect deterministic flows,
    \item maintaining flow continuity when traffic moves between Wi-Fi~6 and 5G domains,
    \item applying PDB constraints to satisfy TSN deadlines by configuring low-PDB 5QI flows to receive preferential scheduling and bounded queueing delays.
\end{itemize}

The Wi-Fi~6 APs interface with the 5G Core through the TSN-enabled Ethernet switch and a core-facing interface. To maintain TSN semantics over Wi-Fi, the model ensures that:
\begin{itemize}
    \item \textbf{PCP tags} from TSN flows are preserved and reflected in Wi-Fi MAC handling,
    \item \textbf{EDCA parameters} (CWmin/CWmax, AIFS, TXOP) are configured to prioritize latency-sensitive traffic,
    \item \textbf{Wi-Fi~6 OFDMA RUs} are dynamically assigned according to flow urgency and network load.
\end{itemize}

Both Wi-Fi~6 and 5G access networks connect to the TSN-aware switch, which performs bridging, VLAN classification, and deterministic forwarding. Node positions and radio coverage are defined using ns-3’s \textit{MobilityHelper}, enabling the study of realistic propagation phenomena such as heterogeneous signal strengths, interference patterns, and overlapping coverage regions.
\section{TSN-Aware Dynamic RU Allocation and Scheduling}
\label{sec:tsn-ru-model}

This section presents the new TSN-aware DynamicRU allocation and scheduling model developed and validated in this work. A unified RU abstraction is adopted for both 5G and Wi-Fi~6 domains, and extended with a dynamic TSN-aware RU allocation algorithm on the Wi-Fi~6 side. The overall objective is to:
\begin{itemize}
    \item prioritize time-sensitive TSN flows (e.g., emergency and control traffic),
    \item maintain low latency and jitter for high-priority streams,
    \item maximise RU utilization in the presence of mixed TSN and best-effort traffic.
\end{itemize}

\subsection{Unified RU-Based Resource Allocation}

RUs are modelled as discrete OFDMA time–frequency blocks that represent the basic schedulable resource in both access domains. To ensure baseline fairness across devices and flows, the model employs a \textit{Round Robin (RR)} scheduling strategy over this unified RU abstraction:
\begin{itemize}
    \item \textbf{5G domain}: RUs are assigned sequentially to UEs so that each device receives a comparable share of time–frequency resources over a complete scheduling cycle.
    \item \textbf{Wi-Fi~6 TSN region}: RUs are alternated among STAs in a similar cyclic fashion, granting each STA predictable and periodic access to OFDMA resources.
\end{itemize}

This RR-based allocation prevents any single device or flow from monopolizing resources and provides a common baseline against which TSN-specific enhancements can be evaluated.

\subsection{Dynamic RU Sharing}

To adapt to fluctuations in traffic load and TSN demands, dynamic resource sharing is introduced on top of the RR policy. The RU set is partitioned into:
\begin{itemize}
    \item \textbf{Exclusive RUs}, reserved for individual devices or flows to guarantee deterministic or latency-sensitive performance,
    \item \textbf{Shared RUs}, available to multiple devices when the number of active flows exceeds the pool of exclusive RUs.
\end{itemize}

Typical examples include:
\begin{itemize}
    \item exclusive assignment of \textbf{RU1} to a high-priority STA or \textbf{RU4} to a 5G UE,
    \item shared allocation of \textbf{RU2} among several STAs or \textbf{RU5} among multiple UEs with similar QoS requirements.
\end{itemize}

The sharing model increases effective system capacity by allowing more active flows than the number of exclusive RUs, at the cost of mild performance degradation the flows involved in sharing. In the ns-3 implementation, the RU mapping is periodically updated based on network conditions by a scheduler function such as \texttt{AllocateRusBasedOnFlows()}, which distributes resources proportionally and cyclically among active devices.

While the RU abstraction is unified, each access domain uses its own scheduler, aligned with the TSN and QoS framework.

\subsection{5G RU Scheduling}
In the 5G RAN, the \textit{ns3::RrFfMacScheduler} cyclically assigns RUs to UEs and enforces QoS via bearer configurations:
\begin{itemize}
    \item \textbf{GBR bearers} (e.g., \textit{GBR\_CONV\_VOICE}) for latency-critical or deterministic flows,
    \item \textbf{non-GBR bearers} for best-effort or elastic traffic.
\end{itemize}

TSN-related flows are mapped to appropriate GBR bearers, ensuring predictable capacity and low delay, while other flows occupy the remaining bandwidth under fair conditions. The scheduler interacts with the 5G Core’s 5QI model, which imposes constraints such as PDB, MBR, and priority level, providing end-to-end behavior consistent with TSN requirements.

\subsection{Wi-Fi~6 RU Scheduling}

On the Wi-Fi~6 side, RU allocation is governed by the OFDMA manager, configured with:
\begin{itemize}
    \item a minimum of one RU per frame,
    \item a maximum of four RUs per frame,
    \item dynamic RU selection enabled.
\end{itemize}
These settings provide the scheduler with flexibility to assign resources each frame, ensuring that high-priority TSN flows can meet latency requirements while maintaining fairness for lower-priority stations.

At each transmission opportunity, the scheduler evaluates the current network state and distributes resources in a way that balances determinism with fairness. High-priority stations are given first access to available RUs, ensuring that their time-sensitive flows meet latency and reliability requirements. When the network becomes congested, lower-priority stations are still accommodated by allowing them to share RUs, preventing starvation while keeping overall utilization balanced. 

The integration of OFDMA-based RU allocation with EDCA-based prioritization allows deterministic TSN flows to achieve reduced access delay and robust performance, while still maintaining fairness for lower-priority users in the Wi-Fi~6 domain.

\subsection{TSN-Aware Dynamic RU Allocation in Wi-Fi~6}

Building on the mechanisms above, the case study introduces a
\textit{flow-aware dynamic RU allocation} procedure at the Wi-Fi~6 AP.
Rather than relying on static, station-centric RU partitioning, the AP
periodically recomputes RU to flow assignments from runtime observations,
allowing the mapping to adapt as flows appear, disappear, or change their
activity levels.

At each scheduling interval (e.g., a Wi-Fi~6 Trigger Frame TxOp), the AP
retrieves per-flow statistics from the ns-3 \textit{FlowMonitor}, including
transmitted and received bytes and the corresponding IPv4 classifier
tuples. Using this information, the AP:

\begin{itemize}
    \item filters inactive flows (with zero Tx/Rx bytes),
    \item discards flows outside the configured valid subnets,
    \item assigns each remaining active flow to an RU using a round-robin
    mapping over the available RU set, and
    \item builds a per-RU sharing table indicating which flows are mapped
    to each RU.
\end{itemize}

This procedure is formalized in Algorithm~\ref{alg:flow_aware_ru}, which
constructs (i) an allocation map $A: f \rightarrow r$ and (ii) a sharing
table $SRU(r)$. After assignment, each RU is labeled as \textit{exclusive}
when it is associated with a single active flow, or \textit{shared} when
multiple flows are mapped to the same RU.

Adaptation arises from the periodic re-execution of this FlowMonitor-driven
mapping: as traffic conditions evolve, the set of active flows changes and
RU assignments are updated accordingly. Under congestion, fair RU sharing
is ensured through the round-robin assignment across flows. QoS
differentiation itself is handled outside the RU mapping via EDCA
configuration and priority translation mechanisms (e.g., DSCP/PCP marking
and TAS scheduling on the wired side), keeping the RU allocation logic
lightweight while enabling end-to-end QoS-aware operation across the
integrated Wi-Fi~6 and TSN (and, where applicable, 5G) network.

\begin{algorithm}[t]
\caption{Flow-Aware Dynamic Resource Unit (RU) Allocation Using FlowMonitor}
\label{alg:flow_aware_ru}
\begin{algorithmic}[1]

\Require
Set of observed IP flows $F$ from FlowMonitor; \\
Per-flow statistics $\{txBytes_f, rxBytes_f\}$; \\
IPv4 classifier tuples $(src_f, dst_f)$; \\
Set of valid IPv4 subnets $\mathcal{N}$; \\
Total number of Resource Units $N_{RU}$

\Ensure
RU allocation map $A : f \rightarrow r$; \\
RU sharing table $SRU$

\State Initialize allocation map $A \leftarrow \emptyset$
\State Initialize RU sharing table $SRU \leftarrow \emptyset$
\State Initialize RU index $r \leftarrow 0$

\State Retrieve flow statistics from FlowMonitor

\For{each flow $f \in F$}
    \If{$txBytes_f = 0$ \textbf{and} $rxBytes_f = 0$}
        \State \textbf{continue} \Comment{Skip inactive flows}
    \EndIf

    \State Extract $(src_f, dst_f)$ using the IPv4 flow classifier

    \If{$src_f \notin \mathcal{N}$ \textbf{or} $dst_f \notin \mathcal{N}$}
        \State \textbf{continue} \Comment{Ignore flows outside valid subnets}
    \EndIf

    \State $r_f \leftarrow r \bmod N_{RU}$ \Comment{Assign RU}
    \State $A(f) \leftarrow r_f$ \Comment{Update allocation map}
    \State $SRU(r_f) \leftarrow SRU(r_f) \cup \{f\}$ \Comment{Update sharing table}
    \State $r \leftarrow r + 1$
\EndFor

\For{each Resource Unit $r \in SRU$}
    \If{$|SRU(r)| > 1$}
        \State Mark RU $r$ as \textit{shared}
    \Else
        \State Mark RU $r$ as \textit{exclusive}
    \EndIf
\EndFor

\State Output RU allocation map $A$ and RU sharing table $SRU$
\State Repeat the procedure periodically at fixed scheduling intervals

\end{algorithmic}
\end{algorithm}

\section{Performance Evaluation and Results}
\label{sec:evaluation}
\label{sec: Evaluation}

This section presents and analyzes the outcomes of two main activities of the study, framed as case study evaluations. The first case study evaluation (\textbf{C1}) examines QoS behaviour across multiple traffic classes in a hybrid Wi-Fi/TSN Ethernet scenario with different combinations of STAs and Ethernet devices. The analysis focuses on throughput, latency, jitter, and packet loss under varying traffic loads, highlighting how EDCA and TAS enforce prioritization in practice. 

A second case study (\textbf{C2}) evaluates the proposed dynamic RU allocation mechanism in Wi-Fi~6, investigating how adaptive resource assignment affects efficiency, fairness, and QoS compared with static allocation. 

\subsection{C1: QoS Traffic Management with EDCA and TAS in a converged wired-wireless network}
C1 evaluates the effectiveness of QoS traffic classification and prioritization in a hybrid TSN WiFi-6 to Ethernet, considering two representative scenarios:
\begin{itemize}
    \item \textbf{Scenario 1:} Two STAs and one ETH device.
    \item \textbf{Scenario 2:} Two STAs and two ETH devices.
\end{itemize}

In all scenarios, wireless connectivity is provided by an AP serving up to two STAs using EDCA. The EDCA mechanism differentiates traffic into access categories, reducing contention for high-priority flows and ensuring that sensitive traffic (such as voice or emergency traffic) obtains expedited access to the wireless medium. On the wired side, one or two Ethernet devices are integrated into a TSN-enabled segment, where a TAS-enabled switch enforces deterministic scheduling and bounded latency. 

Table \ref{tab:trafficflows} defines the traffic classes used in the simulation, including their packet size, data rate, interarrival time, message type, and flow type.

\begin{table}[h]
\centering
\caption{Traffic Class Details and Characteristics.}
\label{tab:trafficflows}
\begin{tabular}{|l|c|c|c|c|c|}
    \hline
    \textbf{Class} & \textbf{Pkt size} & \textbf{Rate} & \textbf{IAT} & \textbf{Type} & \textbf{Intensity} \\ \hline

    Emergency (VO) & 64 B  & 200 kbps & 0.005 s & CBR & Small \\ 
    \multicolumn{6}{|p{7.5cm}|}{\footnotesize Highest-priority traffic with strict latency requirements.} \\ \hline

    Voice (VO) & 128 B & 100 kbps & 0.02 s & CBR & Medium \\ 
    \multicolumn{6}{|p{7.5cm}|}{\footnotesize Real-time voice with low latency but slightly more tolerance than emergency.} \\ \hline

    Video (VI) & 512 B & 500 kbps & 0.033 s & VBR & Large \\ 
    \multicolumn{6}{|p{7.5cm}|}{\footnotesize Video conferencing/streaming with low-latency requirements.} \\ \hline

    Best Effort (BE) & 1024 B & 1 Mbps & 0.1 s & VBR & Large \\ 
    \multicolumn{6}{|p{7.5cm}|}{\footnotesize Non-critical data transfers, no strict delay constraints.} \\ \hline

    Background (BK) & 256 B & 50 kbps & 0.5 s & CBR & Small \\ 
    \multicolumn{6}{|p{7.5cm}|}{\footnotesize Low-priority traffic tolerant of delays.} \\ \hline
\end{tabular}
\end{table}

By applying varying traffic loads and QoS configurations, the study examines how effectively the network prioritizes heterogeneous traffic types and how well QoS mechanisms sustain performance under different operating conditions.

\subsubsection{Scenario 1: Comparative Performance Assessment of EDCA and TAS in a Multi-STA Configuration}
\begin{figure}[h]
    \centering
    \includegraphics[width=\columnwidth]{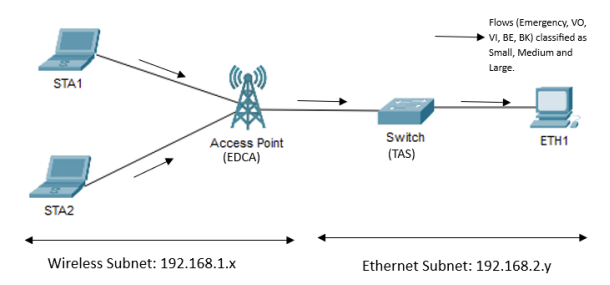}
    \caption{TSN Wi-Fi 6 region with 2 STAs communicating with 1 Ethernet devices in the TSN-Ethernet region.}
    \label{fig:scenario2}
\end{figure}

\begin{figure}[h!]
    \centering
    \includegraphics[width=0.75\linewidth]{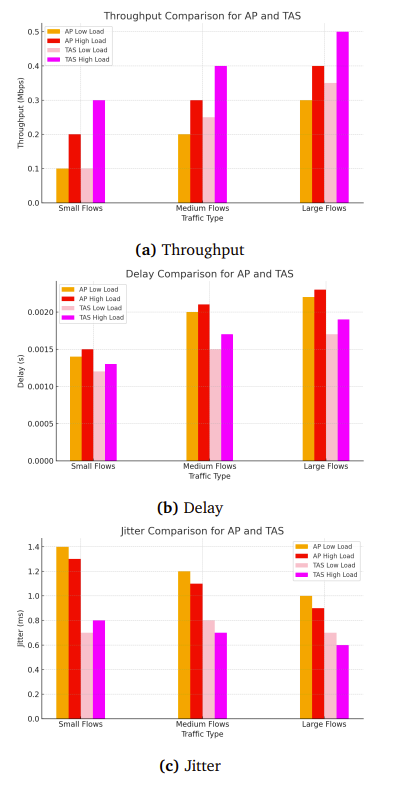}
    \caption{C1 Scenario 1 results, Throughput, Delay, and Jitter.}
    \label{fig:scenario2-results}
\end{figure}

Scenario~1 relies on the topology presented in Fig. \ref{fig:scenario2}. Two STAs communicate with an AP using EDCA, while an Ethernet device exchanges traffic through a TSN-enabled switch operating with TAS. This setup, albeit simple, makes it possible to study how contention-based wireless access interacts with deterministic wired scheduling under varying load conditions. Results are provided in Fig. \ref{fig:scenario2-results}, for throughput, latency and jitter.

The figure provides a comparison on how the AP using EDCA and the TAS-enabled switch behave under both low and high traffic loads across small, medium, and large flows. 

Under low-load conditions, both systems reach similar throughput. However, when the load increases, their behavior diverges. The AP begins to struggle with contention, thus there is a slight reduction in throughput. In contrast, TAS shows improvements, in particular for larger flows under high load, with small flows increasing to approximately 0.30 Mbps, medium flows rising to 0.40 Mbps, and large flows climbing to around 0.55 Mbps. This upward trend reflects TAS’s deterministic scheduling, which scales efficiently even as the traffic variability increases.

Delay results amplify this contrast. Under low load, delay values remain modest for both systems, with small flows reaching around 1.0 ms on the AP and 0.9 ms on TAS, medium flows around 1.4 ms on the AP and 1.0 ms on TAS, and large flows near 2.0 ms on the AP versus 1.7 ms on TAS. 

Under high-load conditions, however, the AP experiences a slight delay increase: small-flow delay rises to roughly 1.3 ms, medium flows reach 1.8 ms, and large flows increase to about 2.3 ms. TAS, by comparison, maintains a slightly similar control over delay, with small flows staying near 1.0 ms, medium flows around 1.2 ms, and large flows approximately 1.9 ms. These results highlight how contention at the AP leads to queue build-up, while TAS’s time-slot-based scheduling limits delay growth.

Jitter is the performance parameter that exhibits more contrast between EDCA and TAS. Under low load conditions, jitter stays moderate on both systems, around 0.80 ms for small flows on EDCA versus 0.60 ms on TAS, and roughly 0.90 ms for large flows on the AP compared with 0.70 ms on TAS. When the load increases, the AP’s jitter rises : small flows reach nearly 1.40 ms, medium flows about 1.20 ms, and large flows around 1.10 ms. In contrast, TAS remains highly stable, with jitter only increasing slightly to about 0.90 ms for small flows, 0.70 ms for medium flows, and 0.55 ms for large flows.  

EDCA’s contention-based access causes higher variability under congestion, while TAS maintains predictable behaviour even at high load.

The results corroborate that while EDCA on the AP performs adequately when traffic loads are low, it degrades under heavy load, especially for small and medium flows. Throughput stagnates, delay and jitter increase. TAS, on the other hand, consistently delivers higher throughput, lower delay, and significantly lower jitter under the same high-load conditions, as expected. This reinforces the strength of deterministic scheduling in maintaining QoS, particularly as network demand increases, and the need for a dynamic RU allocation scheme, that can best support TAS in the TSN Wi-Fi regions.

\subsubsection{Scenario 2: Integrated Wireless-Wired Traffic Interaction}

\begin{figure}[h]
    \centering
    \includegraphics[width=\columnwidth]{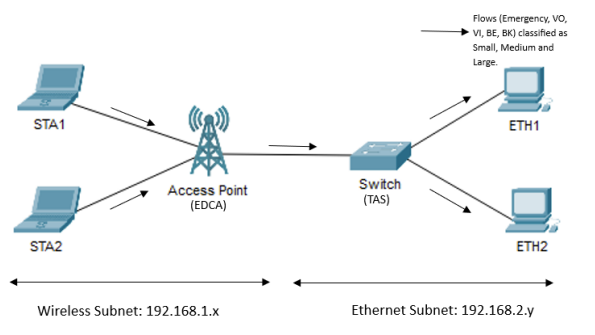}
    \caption{TSN Wi-Fi 6 region with 2 STAs communicating with 2 Ethernet devices in the TSN-Ethernet region.}
    \label{fig:scenario3}
\end{figure}

Scenario 2 provides a simple topology to analyse the behaviour of TAS and EDCA in environments involving more than one STA and more than one Ethernet device, as represented in Fig. \ref{fig:scenario3}. This dumbell topology highlights the AP’s dual-role operation, simultaneously managing contention-based wireless access via EDCA and forwarding traffic into a TSN-enabled wired segment governed by TAS.

\begin{figure}[h!]
    \centering
    \includegraphics[width=0.75\linewidth]{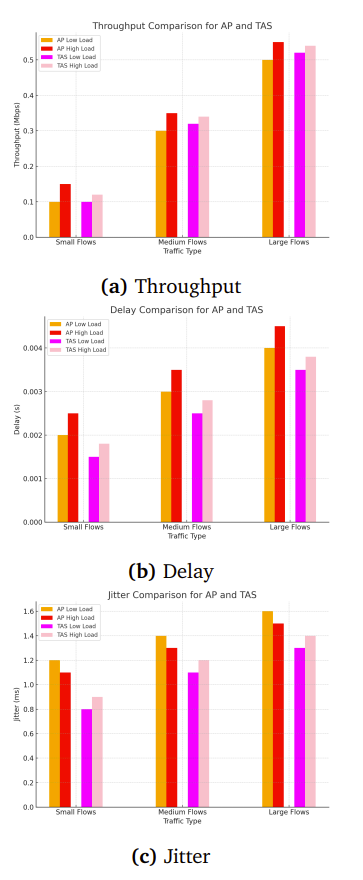}
    \caption{C1 Scenario 2 results, Throughput, Delay, and Jitter.}
    \label{fig:scenario3-results}
\end{figure}

This scenario is designed to study how increasing the number of wireless and wired endpoints impacts end-to-end performance, particularly when heterogeneous traffic types compete for shared resources. The bi-directional packet exchange between all four devices is depicted in Fig. \ref{fig:scenario3-results}, illustrating how traffic flows traverse the wireless and wired domains.

The results again show how EDCA and TAS behave for small, medium, and large flows, under both low-load and high-load network conditions, in terms of achieved throughput, delay, and jitter.

The achieved throughput shows a similar behaviour to the one attained in the previous scenarios, with lower values for EDCA in comparison to TAS.

Under low load, the AP delivers around 0.12 Mbps for small flows and roughly 0.30 Mbps for medium flows, but both values dip when the system becomes congested. In contrast, the TAS-based switch remains far more stable, sustaining about 0.32 Mbps for small flows and 0.42 Mbps for medium flows and 0.55 for large flows.

Delay exhibits similar behaviour patterns, with TAS proportionating less delay, even when the number of flows nearly doubles. This stability reflects TAS’s deterministic, isolation-based scheduling, which prevents queues from expanding uncontrollably the way they do in EDCA.

With EDCA, jitter grows as the AP becomes overloaded: small-flow jitter rises from roughly 1.1 ms to more than 1.4 ms, and medium and large flows also experience noticeable increases as contention intensifies. TAS again remains steady, keeping jitter around 0.8–0.9 ms for small flows and only slightly higher for medium and large flows. Because TAS schedules transmissions in fixed, conflict-free windows, it avoids the unpredictable backoff delays that cause jitter spikes in EDCA.

Altogether, the figure shows that once multiple wireless and wired devices begin competing for TxOPs and channel capacity, EDCA’s contention-based behavior becomes a limiting factor. Throughput drops, delay grows, and jitter becomes difficult to control. By contrast, TAS continues to deliver predictable, high-quality performance across all flow types because its deterministic scheduling isolates traffic classes and prevents congestion from cascading. 

\subsection{C2: Dynamic RU Allocation and Scheduling in a Converged Wired-Wireless Network}

On C2, simulations carried out aim at analysing the performance of the proposed dynamic RU allocation mechanism, focusing on its ability to adapt to varying traffic intensities and comparing its behavior against traditional EDCA-based scheduling. The analysis evaluates how dynamic allocation enhances throughput stability, minimizes packet loss, and maintains service quality by intelligently assigning exclusive or shared RUs as network conditions evolve. Specifically, C2 focuses on understanding how dynamic RU allocation manages network resources under low, medium, and high traffic loads, and how its adaptability and QoS performance compare with EDCA, particularly in scenarios where traffic demands fluctuate sharply.

Dynamic RU allocation provides an adaptive scheduling strategy in which resource distribution automatically adjusts to changes in network demand. This section analyzes its behavior across different traffic conditions, contrasts it with static RU allocation, and highlights potential improvements for future implementations.

\begin{figure}[h!]
    \centering
    \includegraphics[width=1\linewidth]{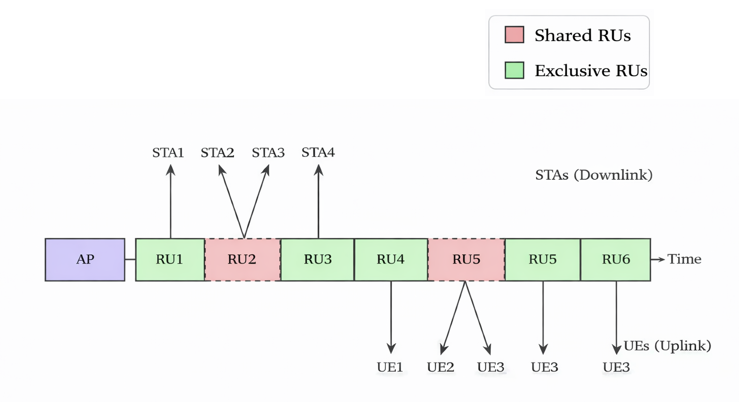}
    \caption{Dynamic RU allocation.}
    \label{fig:RU-allocation}
\end{figure}

Fig. \ref{fig:RU-allocation} illustrates RU allocation in a wireless network, where an AP assigns RUs to STAs (downlink) and UEs (uplink). The AP transmits to STA1-STA4 using RU1-RU3, with RU2 shared between STA2 and STA3, reflecting dynamic allocation. UEs (UE1-UE3) transmit via RU4-RU6, with RU5 shared between UE2 and UE3, optimizing resources.

During the low-traffic period (0–2 seconds), dynamic RU allocation performs with high efficiency by assigning exclusive RUs to active flows. Because each flow receives its own RU, contention is eliminated and the system achieves consistently strong performance. Flow 1 records a throughput of 0.67 Mbps, a delay of 1 ms, and jitter of 0.61 ms, with other flows showing similarly stable behavior. No packet loss or degradation occurs, demonstrating the mechanism’s ability to fully utilize available resources while maintaining low latency and predictable service quality.

As the traffic load increases (3–5 seconds), the system transitions from exclusive to shared RUs, enabling multiple flows to occupy the same resource unit. This shift increases capacity utilization and accommodates more active flows without compromising throughput. For example, Flow 1 still achieves 0.67 Mbps, while Flow 9, which shares RU0, achieves 0.59 Mbps. Although throughput remains stable, increased sharing results in modest rises in delay and jitter: Flow 9 experiences 1.34 ms delay and 0.71 ms jitter. These changes illustrate the system’s balanced approach: it maintains high throughput while absorbing moderate increases in latency as contention naturally emerges.

Under peak traffic (6–9 seconds), all eight RUs are fully utilized and shared among sixteen flows, illustrating the system’s ability to operate under heavy demand. Even with full RU occupancy, no flows are dropped. However, performance becomes more uneven. Flow 8, sharing RU7 with Flow 16, records 0.51 Mbps throughput, 4.16 ms delay, and 1.65 ms jitter, while Flow 16 experiences a lower throughput of 0.336 Mbps. As contention increases, delays and jitter rise, reflecting expected trade-offs at maximum network load. Nonetheless, the system preserves throughput fairness and avoids starvation, showcasing the strong adaptability of dynamic allocation under stress.

\section{Conclusions and Future Work}
\label{sec:conclusion}
The results of this initial study offer an early but meaningful look into how dynamic RU allocation behaves under controlled traffic conditions. Across all scenarios, the mechanism consistently demonstrated an ability to adapt to fluctuating load levels, allocating exclusive RUs when resources were abundant and transitioning to shared use only when demand increased. This flexibility allowed the system to maintain stable throughput even as the network approached saturation, highlighting one of the clearest benefits of dynamic allocation over static, fixed assignment policies.
Under low-traffic conditions, the network operated with high efficiency: each flow received its own RU, eliminating contention and keeping delays and jitter to a minimum. As traffic intensified, dynamic allocation shifted toward shared RUs, distributing resources across more flows without drops or service interruption. Although this introduced expected increases in delay and jitter, the underlying throughput performance remained notably stable. Even at peak load—where all eight RUs were actively shared among sixteen competing flows—the system continued to serve every flow, showing no cases of starvation. This robustness under stress contrasts sharply with the behavior observed in static allocation, where fixed RU assignments quickly led to underutilization at low load and congestion-induced packet losses at high load.

Hence, the study demonstrates that dynamic RU allocation can bring substantial improvements to converged TSN networks, particularly in environments where traffic conditions evolve rapidly. At the same time, the performance trade-offs observed at high loads, especially in latency-sensitive flows underscore the need for further tuning and optimization. 

Because the scenarios studied were intentionally constrained and represent only a subset of realistic operating conditions, the findings should be viewed as exploratory. The work forms a baseline from which future evaluations can expand, incorporating larger network topologies, real-world channel conditions, and more diverse traffic characteristics to validate and generalize the trends identified here.

\def\refname{REFERENCES}
  \bibliographystyle{ieeetr}
  \bibliography{references}

\end{document}